\documentclass[twocolumn,pre,superscriptaddress,bibnotes,notitlepage,nofootinbib]{revtex4-1}
\usepackage{color}
\definecolor{red}{rgb}{0.75,0,0}
\definecolor{blue}{rgb}{0,0,0.75}
\definecolor{green}{rgb}{0,0.5,0}
\usepackage{amsmath}
\usepackage{amssymb}
\usepackage{bm}
\usepackage{mathtools}
\usepackage{graphicx}
\usepackage{hyperref}
\usepackage{verbatim}
\DeclareMathAlphabet{\mathcalligra}{T1}{calligra}{m}{n}

\def\be{\begin{equation}}
\def\ee{\end{equation}}
\def\bea{\begin{eqnarray}}
\def\eea{\end{eqnarray}}

\def\besub{\begin{subequations}}
\def\eesub{\end{subequations}}

\def\bwd{\begin{widetext}}
\def\ewd{\end{widetext}}

\newcommand{\bsf}[1]{\textsf{\textbf{#1}}}

\newcommand{\AM}[1]{\textcolor{black}{#1}}

\begin{document}
\title{Active uniaxially ordered suspensions on disordered substrates}
\author{Ananyo Maitra}
\email{nyomaitra07@gmail.com}
\affiliation{Sorbonne Universit\'{e} and CNRS, Laboratoire Jean Perrin, F-75005, Paris, France}
%\author{Martin Lenz}
%\email{martin.lenz@u-psud.fr}
%\affiliation{LPTMS, CNRS, Univ. Paris-Sud, Universit\'e Paris-Saclay, 91405 Orsay, France}
%\affiliation{MultiScale  Material  Science  for  Energy  and  Environment,
%UMI  3466,  CNRS-MIT,
%77  Massachusetts  Avenue,
%Cambridge,  Massachusetts  02139,
%USA}
\begin{abstract}
Multiple experiments on active systems consider oriented active suspensions on substrates or in chambers confined along one direction. The theories of polar and apolar phases in such geometries were considered in A. Maitra et al, arXiv 1901.01069 and A. Maitra et al., PNAS 115, 6934 (2018) respectively. However, the presence of quenched random disorder due to the substrate cannot be completely eliminated in many experimental contexts possibly masking the predictions from those theories. In this paper, I consider the effect of quenched orientational disorder on the phase behaviour of both polar and apolar suspensions on substrates. I show that polar suspensions have long-range order in two dimensions with anomalous number fluctuations, while their apolar counterparts have only short range order, albeit with a correlation length that can increase with activity, and even more violent number fluctuations than active nematics without quenched disorder. These results should be of value in interpreting experiments on active suspensions on substrates with random disorder. 

\end{abstract}
\maketitle

Biological systems and their artificial analogues convert chemical energy into mechanical work at the scale of \emph{individual} constituents \cite{SRJSTAT, LPDJSTAT}. This breaking of detailed balance at the microscopic scale results in macroscopic stresses and currents \cite{RMP, SRrev, Salbreux, CurieRep, Saintillan} that have been shown to singularly modify the phase behaviour of such ``active systems'' relative to their passive, equilibrium counterparts. This includes the prediction of long range order in two dimensions for motile $XY$ spins on a substrate \cite{TT, TT_rean, TMalthus, TTSR}, the generic destruction of apolar phases in infinite, momentum-conserved systems in any dimension \cite{RMP, Aditi1, Voit} and the anomalous scaling of number fluctuations with the mean number $N$ in orientationally ordered phases which, being larger than $\sqrt{N}$, violates the law of large numbers \cite{Aditi2, TonerGNF}. These predictions have been confirmed in experiments \cite{RMP, Doostmohammadi} and have been shown to be responsible for various biological phenomena ranging from flocking of birds \cite{Cavagna} to spatio-temporally chaotic flows in bacterial fluids \cite{Goldstein} to crawling of cell-layers \cite{celllayer} to spontaneous rotation of the cellular nucleus \cite{Ano_nuc_rot}.

Most experimentally relevant biological active systems and many of their artificial counterparts should be viewed as a suspension of active particles in a fluid medium. Furthermore, many such experiments are conducted in systems which are confined between a substrate and a cover-slip. Boundary conditions fundamentally change the behaviours of active systems \cite{Ano_und, Ano_nuc_rot}. The theories of homogeneous active polar and apolar suspensions on substrates or in confined channels were considered in \cite{Ano_pol} and \cite{Ano_apol} respectively. There, it was shown that contrary to naive expectations, polar fluids in confined channels have long range order due to a singular suppression of polarisation fluctuations due to the long-range effect of fluid incompressibility and have normal (i.e., obeying law of large numbers) number fluctuations \cite{Ano_pol}. It was further demonstrated in \cite{Ano_apol} that the effective elastic modulus of active apolar suspension not only need not become negative (signalling an instability) at any activity, but could also \emph{increase} with activity. An apolar suspension, unlike its polar counterpart has only quasi-long-range order and has giant number fluctuations. However, often in experiments, the substrates and the cover-slips are not perfectly uniform but, due to imperfections, may have local, quenched anisotropies. Such local anisotropies provide a local ``easy-direction'' for the orientation of the active particles. Therefore, it is important to understand how the conclusions regarding order and number fluctuations are modified in the presence of such orientational ``quenched disorder''. It has been known that systems that break a continuous symmetry can not have long-range order below four dimensions in equilibrium \cite{Imry}. The effect of quenched disorder has also been examined in active, motile systems, but \emph{without} a suspending fluid medium, by \cite{Guten1, Guten2, Das} where it was shown that it may be possible for such systems to have quasi-long-range order even in \emph{two} dimensions contrary to expectations borne out of experience with equilibrium systems. 

In this paper, I consider the experimentally relevant case of polar and apolar suspension on substrates (or confined in narrow channels) in the presence of random, quenched orientational disorder. I demonstrate that, within a linearised theory, polar suspensions on substrates have \emph{long-range order} in \emph{two dimensions} due to the stabilisation of orientational fluctuations caused by the non-local constraint of fluid incompressibility. Unlike their counterparts with only annealed noise however \cite{Ano_pol}, the number fluctuations are predicted to be larger than in passive systems and to violate the law of large numbers. In contrast, apolar fluids, which are not motile on average, do not even have quasi-long-range order in two dimensions -- thus, a monodomain nematic phase is not possible even in the presence of infinitesimal random quenched disorder beyond an active Imry-Ma lengthscale. However, this lengthscale is itself predicted to grow with activity signifying that apolar suspensions that are not destabilised at high activities can have large Imry-Ma scales and therefore, at moderate disorder strengths, a monodomain nematic phase may be observed in experiments. Further I show that number fluctuations in regions much smaller than the Imry-Ma length, i.e., at scales where a monodomain nematic exists, are \emph{enormous}, much larger than the corresponding system with annealed noise. I now demonstrate how I obtain these results in detail, first considering a polar suspension with quenched disorder and then an apolar one.

\section{Polar suspensions on disordered substrates}
The equations of motion of a polar suspension on a substrate in the \emph{absence} of quenched disorder have been considered in detail in \cite{Ano_pol}. I adapt these equations to describe a situation in which
%We describe the theory of a polar suspension on a substrate. 
the presence of impurities on the substrate create local ``easy-directions'' for the polarisation field of the suspension ${\bf p}$ which acts as a quenched noise for {\bf p}. The concentration of the active particles is denoted by $c$ and the centre-of mass velocity of the fluid and the active particles by ${\bf u}$. 
%The equations of motion are the same as in \cite{ano_pol}, which we recapitulate for completeness, with the \emph{only} difference being the presence of an additional quenched random noise in the polarisation equation. 
The joint density of the fluid and the active particles together is incompressible leading to the constraint $\nabla\cdot{\bf u}=0$. The dynamical equations are
\begin{equation}
\partial_t{\bf p}+{\bf u}\cdot\nabla{\bf p}+ {\lambda_a}{\bf
p}\cdot\nabla{\bf p}-{\boldsymbol{\omega}}\times{\bf p}= {\Lambda}{\bf
u}-\lambda{\bf p}\cdot{{\bsf U}}-\frac{\delta \mathcal{H}}{\delta {\bf p}} +\bm{\xi}_Q,
\label{p2d}
\end{equation}
where $\bsf{U}$ is the symmetric part of $\nabla{\bf u}$, $\boldsymbol{\omega}=(1/2)\nabla\times{\bf u}$, $\lambda_a$ denotes the strength of active self-advection and $\lambda$ is the flow-alignment parameter. The $\Lambda$ term describes the fact that on a substrate, polarisation responds to the local velocity and not only its gradient \cite{Ano_pol, Harsh, Lauga}. The free-energy $\mathcal{H}$, which would have controlled the statics and dynamics in the absence of activity is 
\begin{equation}
\label{fenergp}
\mathcal{H}=\int_{\bf x}\left[{\alpha(c)\over2}|{\bf p}|^2+{\beta\over4}|{\bf p}|^4+ {K\over2}|\nabla{\bf p}|^2+{\gamma} {\bf p}\cdot\nabla c+c\ln c\right].
\end{equation}
Finally, $\bm{\xi}_Q$ denotes a quenched random force whose statistics I take to be Gaussian \cite{Guten1, Guten2, Das}: 
\begin{equation}
\langle\xi_{Q_i}({\bf x})\xi_{Q_j}({\bf x}')\rangle=T_Q\delta_{ij}\delta({\bf x}-{\bf x}')
\end{equation}
where ${\bf x}$ is the spatial coordinate, $T_Q$ is the strength of the disorder and the angular brackets imply an average over the quenched disorder. 
The equation for the velocity field on a substrate reads 
\begin{multline}
\Gamma {\bf u}= \upsilon{\bf p}-\nabla
 \Pi+\zeta_1{\bf p}\cdot\nabla{\bf p}+\zeta_2{\bf p}\nabla\cdot{\bf p}\\- {\Lambda}\frac{\delta \mathcal{H}}{\delta {\bf p}}-\lambda\nabla\cdot\left[{\bf p}\frac{\delta \mathcal{H}}{\delta {\bf p}}\right]^S-\nabla\cdot\left[{\bf p}\frac{\delta \mathcal{H}}{\delta {\bf p}}\right]^A,
\label{v2d}
\end{multline}
where $\Pi$ enforces the incompressibility condition $\nabla\cdot{\bf u}=0$ and superscripts $S$ and $A$ denote symmetric and antisymmetric parts of a tensor. $\upsilon{\bf p}$ describes the active motility and $\zeta_1$ and $\zeta_2$ are active terms at the next order in gradients \cite{Ano_apol}. The remaining terms are required by Onsager symmetry to obtain an equilibrium system in the limit $\upsilon=\zeta_1=\zeta_2=\lambda_a=0$. Note that I have not considered active and passive higher order in gradients contributions to the force-density explicitly since they are subdominant at long wavelengths to the terms in \eqref{v2d} and \eqref{p2d}.
%In principle, there is another independent active term at second order in gradients -- these are similar to the splay and bend elastic terms that could arise from $\delta\mathcal{H}/\delta {\bf p}$ if we had not written the energy in a single Frank constant approximation.  
The equation for the concentration fluctuations, $\delta c=c-c_0$, where $c_0$ is the mean concentration of active particles, is 
\begin{equation}
\label{coneqn}
\partial_t\delta{c}+{\bf u}\cdot\nabla \delta c=-\nabla \cdot (v_p {c}{\bf p}- D_c \nabla \delta {c}),
\end{equation}
to lowest order in fields and gradients, where the term with the coefficient $\upsilon_p$ describes an active current proportional to the polarisation and $D_c$ is the isotropic diffusivity. I have ignored a further active contribution to the current $\propto ({\bf pp}\cdot{\bf u})$ since it can be shown to not qualitatively affect the fluctuation spectrum of a stable polar phase.
%\AMN{Beside these, the concentration equation can also include further terms that depend on ${\bf u}$, the simplest of which, in this system in which momentum is not conserved, is $\propto\nabla\cdot({\bf pp}\cdot{\bf u})$. We ignore this however since, as we will show later, this cannot qualitatively modify the static structure factor of concentration fluctuations.} 
The impurity field is assumed to \emph{only} affect the local polarisation. There are additional \emph{annealed} noises in these equations, but in this paper I will only consider \emph{equal-time} correlations whose dominant contribution arises from the quenched component (the equal-time correlations in the absence of quenched noise but with annealed noise was calculated in \cite{Ano_pol}). 

As shown in \cite{Ano_pol}, 
%The static, homogeneous, disorder-free linear equations for ${\bf p}$ and ${\bf u}$ are
%\begin{equation}
%0=-\alpha {\bf p}_0-\beta p_0^2{\bf p}_0+\Lambda{\bf u}_0
%\end{equation}
%\begin{equation}
%{\bf u}_0=\frac{\upsilon}{\Gamma} {\bf p}_0-\frac{\Lambda}{\Gamma}\alpha {\bf p}_0-\frac{\Lambda}{\Gamma}\beta p_0^2{\bf p}_0.
%\end{equation}
for $\tilde{\alpha}=\alpha(c_0)-\Lambda\upsilon/(\Gamma+\Lambda^2)=\alpha-w<0$, the disordered state (with $p_0=0$) is unstable and an ordered state with $p_0^2=\sqrt{\tilde{\alpha}/\beta}$ sets in. I take the direction of ordering to be along $\hat{x}$. The uniform fluid velocity ${\bf u}_0=(w/\Lambda)p_0\hat{x}$ in this state \cite{Ano_pol}. I now consider the statistics of small fluctuations of the polarisation $\delta{\bf p}=(p_0+\delta p)(\cos\theta\hat{x}+\sin\theta\hat{y})-p_0\hat{x}$ and concentration $\delta c$ about this state in the presence of quenched disorder. Since it was shown in \cite{Ano_pol} that a stable aligned phase is only possible for $w>0$, this is the case I will consider. Fourier transforming in space and eliminating the incompressible velocity field and the fast $\delta p$ field, I obtain the coupled, linearised equation for $\theta_q$ and $\delta c_q$. Only retaining up to $\mathcal{O}(q)$ terms in the $\theta_q$ equation (since, this will be sufficient for calculating the static structure factor in the limit of small wavevectors), I obtain
\begin{equation}
\partial_t \theta_q=-i\frac{\gamma}{p_0} q_y\delta{c}-i  {\Lambda}_a q_x\theta
-w\frac{q_y^2}{q^2}\theta_q+\frac{1}{p_0}\xi_Q
\label{polang}
\end{equation}
where the expression for $\Lambda_a$ is given in the appendix,
%\begin{equation}
%{\Lambda}_a(\phi)=\left(\lambda_a+\frac{w}{\Lambda}\right)p_0-\frac{w p_0}{2\Lambda}\left(1-\lambda\frac{q_x^2-q_y^2}{q^2}\right)-\frac{\Lambda p_0^2}{\Gamma}\left[\left(\zeta_1+w\frac{1-\lambda}{2}\right)\frac{q_x^2}{q^2}-\left(\zeta_2-{w}\frac{1+\lambda}{2}\right)\frac{q_y^2}{q^2}\right]
%\end{equation}
and $\xi_Q$ is the projection of $\bm{\xi}_Q$ transverse to ${\bf p}$. The concentration equation is
\begin{equation}
\partial_t \delta{c}_q=-D_cq^2\delta{c}-i v_p{c}_0p_0q_y\theta-i
\left(v_p+\frac{w}{\Lambda}\right)p_0q_x\delta{c}
\label{concp}
\end{equation}
%\AMN{We note that at this stage it becomes clear that including a term of the form $\nabla\cdot{\bf pp}\cdot{\bf u}$ would not have qualitatively changed the concentration fluctuations. This would have led to a term $\propto q_x^2q_y/q^2\theta$ and since $q_x^2q_y/q^2<q_y$ everywhere, this would not have led to a qualitatively different anisotropy of the static structure facto of concentration fluctuations.} 
The steady state solution of \eqref{polang} and \eqref{concp} are simply obtained by setting $\partial_t\theta_q=\partial_t\delta c_q=0$. 
%and solving the matrix equation
%\begin{equation}
%\begin{pmatrix}i  {\Lambda}_a q_x
%+w\frac{q_y^2}{q^2} && i\frac{\gamma}{p_0} q_y\\i v_p{c}_0p_0q_y && i
%\left(v_p+\frac{w}{\Lambda}\right)p_0q_x+D_cq^2\end{pmatrix}\begin{pmatrix}\theta_q\\\delta c_q\end{pmatrix}=\begin{pmatrix}\frac{1}{p_0}\xi_Q\\0\end{pmatrix}
%\end{equation}
The disorder-averaged correlation function for angular fluctuation can be calculated straightforwardly from the solution of the steady state equations, which in the limit of small wavevectors is 
%The solution of this equation then straightforwardly yields the disorder-averaged correlation functions for $\theta_q$ and $\delta c_q$:
%\begin{equation}
%\langle|\theta_q|^2\rangle=\frac{T_Qq^4[D_c^2q^4\Lambda^2+p_0^2q_x^2(w+\upsilon_p\Lambda)^2]}{-2 c_0p_0q^4q_x^2q_y^2\upsilon_p\gamma\Lambda(w+\upsilon_p\Lambda)\Lambda_a+p_0^2q_x^2(w+\upsilon_p\Lambda)^2(q_y^4w^2+q^4q_x^2\Lambda_a^2)+q^4\Lambda^2[q_y^4(D_cw+c_0\upsilon_p\gamma)^2+D_c^2q^4q_x^2\Lambda_a^2]}
%\end{equation}
%\begin{equation}
%\langle|\theta_q|^2\rangle=\frac{T_Q(w+\upsilon_p\Lambda)^2}{-2 c_0p_0q_y^2\upsilon_p\gamma\Lambda(w+\upsilon_p\Lambda)\Lambda_a+p_0^2(w+\upsilon_p\Lambda)^2[(q_y^4/q^4)w^2+q_x^2\Lambda_a^2]+(q_y^4/q_x^2)\Lambda^2(D_cw+c_0\upsilon_p\gamma)^2}
%\end{equation}
\begin{equation}
\label{ssfang}
\langle|\theta_q|^2\rangle=\frac{T_Q}{p_0^2[(q_y^4/q^4)w^2+q_x^2\Lambda_a^2]}\approx \frac{T_Qq^4}{p_0^2(q_y^4w^2+q_x^6\Lambda_a^2)}.
\end{equation}
This implies that $\langle|\theta_q|^2\rangle\propto1/q^2$ for $q_y^4\lesssim q_x^6$, i.e., largest for $q_y^4\sim q_x^6\implies q_y\sim q_x^{3/2}$. Note that this is somewhat different from the equal-time correlator in the presence of \emph{only} annealed noise \cite{Ano_pol}: $\sim q_x^2/(w q_y^2+K q_x^4)$.
However, it is clear that the real space angular fluctuations $\langle\theta({\bf x})^2\rangle=\int d^2q/(2\pi)^2 \langle|\theta_q|^2\rangle$ converges in the infrared, $\sim1/\sqrt{L}$ as the system size $L\to\infty$ (again, in contrast to the case with annealed noise where it scaled as $\sim 1/L$), signifying the presence of long-range order in two dimensions. More formally, by examining the behaviour of the equal-time correlator under the transformation $x\to b x$, $y\to b^\zeta y$, $t\to b^z t$, $\theta\to b^\chi\theta$, I can calculate the linear roughness $\chi$, anisotropy $\zeta$ and dynamical $z$ exponents. Since the linear fluctuations are governed by $\Lambda_a$ (which can be replaced by its isotropic value for this argument), $w$ and $T_Q$, choosing exponent values such that these quantities remain invariant under the rescaling transformation will ensure that $\langle|\theta_q|^2\rangle$ remains unchanged. Under rescaling, $\Lambda_a\to b^{z-1}\Lambda_a$, $w\to b^{z-2\zeta+2} w$ and $T_Q\to b^{(2z-2\chi-\zeta-1)}T_Q$, which implies that $z=1$, $\zeta=3/2$ and $\chi=-1/4$. The linear anisotropy exponent is consistent with the fact that the fluctuations are strongest in the regime $q_y\sim q_x^{3/2}$ and the roughness exponent implies that the real space angular fluctuations $\langle\theta({\bf x})^2\rangle\sim L^{2\chi}$ decay as $1/\sqrt{L}$ for $L\to\infty$, as discussed earlier. The negativity of the roughness exponent thus implies that the polar, moving state has long range order despite the presence of disorder.

%Integrating this over $q_y$ from $-\infty$ to $\infty$ and over $q_x$ from $-1/L$ to $1/L$, we see that $\langle\theta^2({\bf x})\rangle=\int_{1/L}dq_x\int dq_y(1/2\pi)^2\langle|\theta|^2\rangle\sim 1/\sqrt{L}$ i.e., it vanishes at large distances. More formally, defining the rescaling $x\to bx$, $y\to b^\zeta y$ and $\theta\to b^\chi\theta$, such that the dominant regime of wavevectors is $q_y\sim b^\zeta q_x$, and noting that the dominant regime is given by $q_y^4\sim q_x^6\implies q_y\sim q_x^{3/2}$, implies that the linear anisotropy exponent is  $\zeta=3/2$ (in contrast to the case with annealed noise, where it was $2$). Similarly, since $\langle\theta^2({\bf x})\rangle\sim L^{-1/2}$, $\chi=-1/4$ (again, in contrast to the case with annealed noise where it was $-1/2$).

To calculate the number fluctuations, I now consider the concentration fluctuations
%\begin{equation}
%\langle|\delta c_q|^2\rangle=\frac{T_Qc_0^2q^4q_y^2\upsilon_p^2\Lambda^2}{-2c_0p_0q^4q_x^2q_y^2\upsilon_p\gamma\Lambda(w+\upsilon_p\Lambda)\Lambda_a+p_0^2q_x^2(w+\upsilon_p\Lambda)^2(q_y^4w^2+q^4q_x^2\Lambda_a^2)+q^4\Lambda^2[q_y^4(D_cw+c_0\upsilon_p\gamma)^2+D_c^2q^4q_x^2\Lambda_a^2]}
%\end{equation}
\begin{widetext}
\begin{equation}
\langle|\delta c_q|^2\rangle=\frac{T_Qc_0^2q_y^2\upsilon_p^2\Lambda^2}{-2c_0p_0q_x^2q_y^2\upsilon_p\gamma\Lambda(w+\upsilon_p\Lambda)\Lambda_a+p_0^2q_x^2(w+\upsilon_p\Lambda)^2[(q_y^4/q^4)w^2+q_x^2\Lambda_a^2]+q_y^4\Lambda^2(D_cw+c_0\upsilon_p\gamma)^2}
\end{equation}
\end{widetext}
\AM{This goes as $q^0$ for most directions of the wavevector space. However, for $q_x\lesssim q_y^2$, (i.e., around ${\bf q}\approx q_y\hat{y}$), it diverges as $\sim 1/q_y^2$ and has another peak at around $q_y^2\sim q_x^3$, where it diverges as $1/q_x$. The strongest divergence is for wavevectors ${\bf q}\approx q_y\hat{y}$ whose contribution must dominate the integral of  $\langle|\delta c_q|^2\rangle$ over all wavevectors. In the direction $q_x\lesssim q_y^2$, concentration is the only hydrodynamic field, with a linear diffusive equation of motion and driven by a conserving but quenched noise. In this regime, the angular fluctuations that enter the concentration equation \eqref{concp} can be simply replaced by $\theta=\xi_Q/p_0w-(i\gamma/p_0w)q_y\delta c$. The first leads to the quenched conserved noise, while the second just modifies the diffusivity, implying a $1/q^2$ divergence of the static structure factor in this regime of the wavevector space.
%(for ${\bf q}\approx q_x\hat{x}$, it becomes $\sim q_y^2/q_x^4\ll 1$ since $q_y^2\ll q_x^4$ in this limit). 
%and around ${\bf q}\approx q_x\hat{x}$ where it diverges as $\sim q_x^2/q_y^2\sim 1/q_y^2$.
For $q\approx q_y$, $\langle|\delta c_q|^2\rangle$ may be approximated as 
\begin{equation}
\langle|\delta c_q|^2\rangle\approx\frac{T_Qc_0^2\upsilon_p^2\Lambda^2}{p_0^2\delta\phi^2(w+\upsilon_p\Lambda)^2w+q^2\Lambda^2(D_cw+c_0\upsilon_p\gamma)^2}
\end{equation}
where $\delta\phi$ is the angle between ${\bf q}$ and $\hat{y}$ (note the unusual definition). Now, performing an integral over the angular variable by noting that this integral will be dominated by the contribution around $\delta\phi\approx 0$ and therefore, the range of the integral can be extended to infinity, I obtain $\langle|\delta c_q|^2\rangle\sim{1}/{q}$. This $1/q$ divergence of the static structure factor of concentration fluctuations implies that the R.M.S. number fluctuations $\sqrt{\langle\delta N^2\rangle}$ in a region of size $\ell$ scales as $\langle N\rangle ^{3/4}$, where $\langle N\rangle$ is the mean number of particles in the region, instead of $\sqrt{\langle N\rangle}$, as it would in equilibrium systems not at a critical point \cite{RMP}. Further, such anomalous number fluctuation are also predicted to be \emph{absent} in polar suspensions in the \emph{absence} of quenched disorder \cite{Ano_pol} where $\sqrt{\langle\delta N^2\rangle}$ scales as $\sqrt{\langle N\rangle}$ as in passive systems.
%\begin{equation}
%\langle|\delta c_q|^2\rangle\approx\int^\infty_{-\infty}\delta\phi\frac{T_Qc_0^2\upsilon_p^2\Lambda^2}{p_0^2\delta\phi^2(w+\upsilon_p\Lambda)^2w+q^2\Lambda^2(D_cw+c_0\upsilon_p\gamma)^2}\sim\frac{1}{q}
%\end{equation}
%This small wavevector divergence of the static structure factor of concentration fluctuations can be rationalised as follows: in the direction $q_x^2\lesssim q_y$, concentration is the only hydrodynamic field, with a linear diffusive equation of motion and driven by a conserving but quenched noise. In this regime, the angular fluctuations that enter the concentration equation \eqref{concp} can be simply replaced by $\theta=\xi_Q/p_0w-(i\gamma/p_0w)q_y\delta c$. The first leads to the quenched conserved noise, while the second just modifies the diffusivity, straightforwardly implying the $1/q^2$ divergence for $q\approx q_y$.
Furthermore, the concentration dynamics is unaffected by any nonlinearity in this regime because 
%that the concentration dynamics in this regime (the angular dynamics is massive) is simply given by the linear diffusion equation with a quenched conserved noise can be checked by noting that 
the allowed nonlinearities $\partial_x\delta c^2\ll \partial_y^2\delta c^2$, since this is the regime of the wavevector space under consideration, and $\partial_y^2\delta c^2\ll\partial_y^2\delta c$ since a $1/q$ divergence of the concentration static structure factor implies that the real-space concentration fluctuations vanish at large scales as $1/L$. }

The conclusions reached on the basis of the linear theory may, however, be modified by relevant nonlinearities. I have already argued above that the dynamics is essentially linear for ${\bf q}\approx q_y\hat{y}$. 
In the regime $q_y\sim q_x^{3/2}$, where angular fluctuations are the largest and concentration fluctuations are also divergent at small $q$, albeit not as strongly as along $\hat{y}$, the size of linear concentration fluctuations is determined by $\upsilon_p$ (the choice of the linear dynamical, anisotropy and roughness exponents fixes the other quantities). Therefore,  additionally rescaling $\delta c\to b^{\chi_c}\delta c$ and demanding that the concentration fluctuations remain invariant implies $\upsilon_p\to b^{(z-\chi_c+\chi-\zeta)}\upsilon_p$ must be held fixed. This yields the linear exponent  $\chi_c=-3/4$ ($\chi_c$ is smaller than $-1/2$ -- the value which would have been expected based on $\int_{\bf q}\langle|\delta c_q|^2\rangle\sim 1/L$ -- because for $q_y\sim q_x^{3/2}$, the small wavevector divergence of $\langle|\delta c_q|^2\rangle$ is weaker than for $q\approx q_y$). 
%One may have expected $\chi_c$ to be $-1/2$ since I have argued that $\langle|\delta c_q|^2\rangle\sim 1/q$ which implies $\int_{\bf q}\langle|\delta c_q|^2\rangle\sim 1/L$. However, note that the regime $q_y\sim q_x^{3/2}$ explicitly excludes $q\approx q_y$ which has the largest contribution to concentration fluctuations. 
As in the corresponding annealed noise problem \cite{Ano_pol, CLT} and as discussed in the appendix, the nonlinearities in the $\theta_q$ equation have the forms $(q_y/q_x)(\theta^2)_q$, $(\theta^3)_q$ (both with coefficient $w$), $q_y(\theta^2)_q$, $\delta c_{q-k} k_x\theta_k$ and $q_y(\delta c^2)_q$, while those in the concentration equation have the form $q_x(\delta c^2)_q$, $q_y\theta_{q-k}\delta c_k$ and $ q_x(\theta^2)_q$. Using the linear exponents obtained above, I find that the only relevant nonlinearities in the $\theta_q$ equation are $(q_y/q_x)(\theta^2)_q$ and $ (\theta^3)_q$, both of which must rescale the same way as the linear $(q_y^2/q^2)\theta_q$ term, while concentration equation has only one relevant nonlinearity $ q_x(\theta^2)_q$. In the annealed noise model, if one ignored the concentration fluctuations, one could obtain the exact static exponents \cite{Ano_pol, CLT} through a mapping to a two-dimensional smectic, and ultimately, to a $1+1$ dimensional KPZ equation \cite{Kashuba,Golubovic}. However, an equivalent mapping is not available here. Nevertheless, these nonlinearities must change the linear exponents discussed earlier. This can be seen by a simple argument: within the linear theory, the nonlinearities $(q_y/q_x)(\theta^2)_q$ and $ (\theta^3)_q$ grow, while the linear term $(q_y^2/q^2)\theta_q$ does not. However, due to rotation invariance, the coefficient of all three terms must rescale the same way (which implies $\chi=1-\zeta$) implying that the exponents calculated above have to be modified by fluctuations. This parallels the situation with only annealed noise \cite{Ano_pol, CLT}, where the linear exponents ($\chi=-1/2$, $\zeta=2$) do not satisfy this relationship, but the exact exponents ($\chi=-1/2$, which remains unchanged and $\zeta=3/2$) do. While I have not been able to calculate the exact exponents for this model, it is reasonable to assume that nonlinearities will not destroy the long-range order in this system with quenched disorder as well. If that is the case, the concentration fluctuations are bound to remain anomalous, as predicted by the linear theory.

\section{Apolar suspensions on disordered substrates}
I now contrast the behaviour of polar suspensions on substrates with quenched disorder with apolar suspensions. The deterministic parts of the equations of motion for the apolar suspension will be the same as in \cite{Ano_apol}. I denote angular fluctuations about a homogeneous state, with concentration $c_0$, ordered nematically along the $\hat{x}$, by the angle field $\theta({\bf x})$, the concentration fluctuations by $\delta c=c-c_0$, and an overdamped velocity field by ${\bf u}$. As in the polar model, I will ignore time-dependent, annealed fluctuations since their contribution to the static correlations are subdominant to the quenched disorder considered here. The equation of motion of the angular fluctuations is 
% The linear theory of apolar suspension on substrate with an annealed noise was discussed in \cite{ano_apol}. Here we ask how the conclusions reached in that calculation are modified if the substrate is dirty, leading to a local quenched disorder field for the director dynamics. We consider the stability of a nematic film ordered along the $\hat{x}$ direction to the presence of quenched disorder.
\begin{equation}
\label{orient}
\dot{\theta}=\frac{1-\lambda}{2}\partial_xu_y-\frac{1+\lambda}{2}\partial_yu_x-\Gamma_\theta\frac{\delta  \mathcal{H}}{\delta\theta}+\xi_Q,
\end{equation}
where $|\lambda|>1$ describes particles with a tendency to align under a shear flow, $\xi_Q$ is a quenched orientational noise which has the correlation $\langle\xi_{Q}({\bf x})\xi_{Q}({\bf x}')\rangle=T_Q\delta({\bf x}-{\bf x}')$ as in the polar system, and
%\begin{equation}
%\langle\xi_{Q}({\bf x})\xi_{Q}({\bf x}')\rangle=T_Q\delta({\bf x}-{\bf x}')
%\end{equation}
%and
\begin{equation}
\label{eq:freeenergy}
\mathcal{H}=\int d^2{\bf r} \left[\frac{K}{2}(\nabla \theta)^2+\gamma c\partial_x\partial_y\theta+c\ln c\right],
\end{equation}
where $K>0$ characterizes the tendency of the particles to align.
The adsorbed equation for the flow velocity is
\begin{equation}\label{eq:Darcy}
\Gamma \mathbf{u}=-\nabla\Pi+\mathbf{f}^p+\mathbf{f}^a,
\end{equation}
where $\Gamma$ is the friction coefficient against the substrate, the {pressure $\Pi$ serves as a Lagrange multiplier enforcing the incompressibility condition $\nabla\cdot\mathbf{u}=0$ for the suspension as a whole, while still permitting fluctuations in the concentration of suspended particles.} Onsager symmetry and Eq.~(\ref{orient}) yield the density of passive (equilibrium) forces
\begin{equation}
{\bf f}^p=-\frac{1+\lambda}{2}\partial_y\left(\frac{\delta  \mathcal{H}}{\delta\theta}\right)\mathbf{\hat{x}}+\frac{1-\lambda}{2}\partial_x\left(\frac{\delta  \mathcal{H}}{\delta\theta}\right)\mathbf{\hat{y}}.
\end{equation}
while, the active force density $\mathbf{f}^a$, to lowest order in gradients is
\begin{equation}\label{eq:activeforce}
{\bf f}^a=-(\zeta^Q_1\Delta\mu+\zeta^Q_2\Delta\mu)\partial_y\theta\hat{x}-(\zeta^Q_1\Delta\mu-\zeta^Q_2\Delta\mu)\partial_x\theta\hat{y}
\end{equation}
where $\zeta^Q_1$ and $\zeta^Q_2$ are two independent phenomenological constants, and $\Delta\mu$ denotes the strength of the overall activity in the system.
The evolution of the concentration fluctuations $\delta c$ is governed by a conservation equation 
\begin{equation}
\label{conc}
\partial_t \delta c=D_c\nabla^2\frac{\delta \mathcal{H}}{\delta c}+\zeta_c\Delta\mu\partial_x\partial_y\theta=D_c\nabla^2\delta c+\zeta_c\Delta\mu\partial_x\partial_y\theta
\end{equation}
where the active term $\zeta_c\Delta\mu$ couples orientation fluctuations with concentration fluctuations, and is a standard feature of active nematics~\cite{Aditi2, RMP}. Upon eliminating the fluid velocity ${\bf u}$, the Fourier-transformed equation for angular fluctuations become
\begin{widetext}
\begin{equation}
\label{ang}
\partial_t\theta_q=-q^2\left[\frac{\Delta\mu \{q^2-\lambda(q_x^2-q_y^2)\}\{\zeta^Q_2q^2-\zeta^Q_1(q_x^2-q_y^2)\}}{2\Gamma q^4}+\Gamma_\theta K\right]\theta_q+q_xq_y{\Gamma_\theta\gamma}\delta c_q+\xi_Q=-q^2\bar{K}\theta_q+q_xq_y\Gamma_\theta\gamma\delta c_q+\xi_Q
\end{equation}
\end{widetext}
where $\bar{K}$ is the nematic stiffness renormalised by activity. It was shown in \cite{Ano_apol} that $\bar{K}$ can be positive even at arbitrarily high $\Delta\mu$ when $\zeta_2>|\zeta_1|$ and $|\lambda|<1$ \cite{Ano_apol}. A negative $\bar{K}$ would signal a linear instability of the ordered state. Here, I concentrate on $\bar{K}>0$.
%, $\phi$ is the angle between the wavevector direction and $\hat{x}$ and
%\begin{equation}
%K(\phi)=K_1\cos^2\phi+K_2\sin^2\phi.
%\end{equation}
%\begin{equation}
%\label{concq}
%\partial_t\delta c_q=-q^2D_c\delta c_q-\zeta_c\Delta\mu q_xq_y\theta_q
%\end{equation}
The static correlators for angular and concentration fluctuations are calculated from \eqref{ang} and the Fourier-transformed version of \eqref{conc} after setting $\partial_t\theta=\partial_t\delta c=0$.
%Following the polar calculation, we now calculate the equal-time correlators of angular and concentration fluctuations by solving the steady-state $\partial_t\theta_q=\partial_t\delta c_q=0$ equations 
%\begin{equation}
%\begin{pmatrix}\bar{K}q^2 && -q_xq_y\Gamma_\theta\gamma\\\zeta_c\Delta\mu q_xq_y && D_c q^2\end{pmatrix}\begin{pmatrix}\theta_q\\\delta c_q\end{pmatrix}=\begin{pmatrix}\xi_Q\\0\end{pmatrix}
%\end{equation}
and averaging over the quenched noise. The static structure factor of angular correlations, averaged over the quenched disorder is
{\begin{equation}
\label{apolangfluc}
\langle|\theta_q|^2\rangle=\frac{T_QD_c^2q^4}{(D_c\bar{K}q^4+q_x^2q_y^2\gamma\Gamma_\theta\Delta\mu\zeta_c)^2}\propto\frac{1}{q^4}.
\end{equation}
This implies that the angular fluctuations diverge at least as strongly as $1/q^4$ in \emph{all} directions of the wavevector space. If $\gamma\zeta_c<0$, there is a possibility that along some special directions the denominator goes to $0$, but in that case higher order in wavevector terms have to be retained in the equations of motion and the divergence of angular fluctuations at small wavevectors would be \emph{stronger} than $1/q^4$. This implies that the real space  angular correlation function diverges strongly (as $L^2$) with the system size $L$ implying that nematic order is impossible on disordered substrates even for active suspensions (i.e., they only have short-ranged order) just like their passive counterparts. However, the active analogue of the Imry-Ma lengthscale $\xi_{IM}$ scales with the activity $\Delta\mu$, for large activity. To see this, note (as pointed out in \cite{Ano_apol}) that for large $\Delta\mu$, the active term in $\bar{K}$, $\propto\Delta\mu$ dominates the passive one $\propto \Gamma_\theta K$, $\bar{K}\sim\Delta\mu$. Thus, the denominator of \eqref{apolangfluc} $\sim\Delta\mu^2$. The R.M.S. angular fluctuations can be calculated as
\begin{equation}
\langle\theta({\bf x})^2\rangle=\int_{|{\bf q}|\geq1/L}\frac{d{\bf q}}{4\pi^2}\langle|\theta_q|^2\rangle\sim \frac{T_Q}{\Delta\mu^2}L^2
\end{equation}
where $L$ is the system size. Setting the distortion to be $\mathcal{O}(1)$, I obtain the Imry-Ma lengthscale
$
\xi_{IM}\sim{\Delta\mu}/{\sqrt{T_Q}}.
$
This implies that the domain size beyond which the active nematic (which is linearly stable) loses order \emph{increases} with increasing activity, which is another manifestation of active stabilisation discussed in \cite{Ano_apol}. Thus, it is possible to experimentally tune activity to obtain arbitrarily large nematic domains. 

The concentration fluctuations also scale as $\sim1/q^4$ in all directions for lengthscales below $\xi_{IM}$ (above this scale, the nematic state itself is destroyed and a description in terms of the angle field is invalidated).
\begin{equation}
\label{apolconcfluc}
\langle|\delta c_q|^2\rangle=\frac{T_Qq_x^2q_y^2\Delta\mu^2\zeta_c^2}{(D_c\bar{K}q^4+q_x^2q_y^2\gamma\Gamma_\theta\Delta\mu\zeta_c)^2}
\end{equation}
}{This can be used to calculate the mean-squared number fluctuations $\langle\delta N^2\rangle=\langle N^2\rangle-\langle N\rangle^2$ in a box of size $\ell<\xi_{IM}$
%in a box of size $\xi_{IM}$, which scales as the square of the box size. The mean-squared number fluctuations $\langle\delta N^2\rangle=\langle N^2\rangle-\langle N\rangle^2$ in a box of size $\ell<\xi_{IM}$ can be calculated as 
$
\langle \delta N^2\rangle=\int_{\ell\times\ell}d^2 rd^2r'\langle\delta c({\bf r})\delta c({\bf r}')\rangle.
$
Since $\langle\delta c({\bf r})\delta c({\bf r}')\rangle\sim |{\bf r}-{\bf r}'|^2$ in all directions, below the Imry-Ma lengthscale
$
\langle \delta N^2\rangle\sim \ell^{2+4}\implies \sqrt{\langle \delta N^2\rangle}\sim \ell^3.
$
Since $\langle N\rangle\sim \ell^2$, this implies $\langle \delta N^2\rangle\sim \langle N\rangle ^{3/2}$. This implies that if one measures the number fluctuations in active nematic suspensions on a dirty substrate, in a region in which the nematic is ordered (i.e., in a region smaller than the Imry-Ma lengthscale), the number fluctuations are expected to even larger than in active nematics on a clean substrate. Beyond the Imry-Ma scale, the nematic order itself is destroyed and the number fluctuations should be normal and obey the law of large numbers.}

\section{conclusions}
In this paper, I have demonstrated that active polar fluids can have long-range order with anomalous number fluctuations even in the presence of weak disorder. The presence of long-range order in a two-dimensional system with disorder is a further consequence of the active Anderson-Higgs mechanism \cite{Ander,Higg} discussed in detail in \cite{Ano_pol} where it was shown that due to the interplay of activity and the constraint of incompressibility, the Nambu-Goldstone mode \cite{Nambu, Goldstone} associated with the broken rotation symmetry becomes gapped, with the gap vanishing only precisely along the ordering direction. The calculation here suggests that the long-range ordered state should be observed in biological experiments, such as in crawling cell layers \cite{celllayer} in which a certain amount of disorder may be expected. Furthermore, the theory described here should be quantitatively testable even in weakly compressible systems such as the one composed of motile rods in a bath of immotile beads \cite{Harsh} since, as shown in \cite{Ano_pol}, such systems behave as incompressible two-dimensional polar suspensions up to large length scales. In this artificial system, the strength of quenched disorder may be tuned by fixing a subset of the polar particles in space.
It is interesting to note that in \cite{Dauchot2} anomalous number fluctuations are observed in a polar suspension with long-range order, which could, perhaps indicate the presence of quenched disorder. However, the presence of Toner-Tu waves \cite{TT} in that experiment, which should be damped in the kind of incompressible system I consider here, probably indicates that the constraint of incompressibility is not experimentally relevant there due to the large chamber thickness and, instead it should be modelled as a \emph{compressible} polar flock. 
%However, our polar theory here should be testable in 

The theory for the apolar suspensions on substrates should also be testable in a variety of biological and non-biological experiments. First, though I constructed a theory of apolar suspensions, the orientational and density fluctuations should have the same form in \emph{dry} active nematics without a suspending fluid medium, unlike in the polar system. In this context, the prediction of increasing Imry-Ma scale with activity as well as number fluctuations which are \emph{even larger} than in disorder-free active nematics can be tested by introducing a low density of fixed, randomly oriented apolar rods in the experiment described in \cite{Vijay}. Turning now to experiments on bacteria,
a highly ordered nematic state was observed in a suspension of E. coli \cite{Nishiguchi}. Introduction of disorder would imply that the quasi-long-range nematic state observed in that experiment would become short range with a Imry-Ma scale that increases with bacterial swimming speed, which should be proportional to activity. Finally, the predictions of this paper can also be tested in experiments on living liquid crystal, composed of swimming bacteria in a passive, ordered nematic fluid \cite{LLC} by introducing a small degree of quenched disorder in the passive nematic. 

\acknowledgements{I thank Sriram Ramaswamy and Martin Lenz for illuminating discussions.}
%\begin{acknowledgments}
%I thank Sriram Ramaswamy and Martin Lenz for illuminating discussions.
%\end{acknowledgments}

% cannot have q order. However, we predict that 
\onecolumngrid
\appendix*
\section{Polar suspension with quenched disorder details}
%\subsection{Polar suspensions}
As discussed in the main text, the Fourier transformed linear equations of motion for a polar suspension in the presence of disorder are 
\begin{equation}
\partial_t \theta_q=-i\frac{\gamma}{p_0} q_y\delta{c}-i  {\Lambda}_a q_x\theta
-w\frac{q_y^2}{q^2}\theta_q+\frac{1}{p_0}\xi_Q
\label{polanga}
\end{equation}
where
%\begin{widetext}
\begin{equation}
{\Lambda}_a=\left(\lambda_a+\frac{w}{\Lambda}\right)p_0-\frac{w p_0}{2\Lambda}\left(1-\lambda\frac{q_x^2-q_y^2}{q^2}\right)-\frac{\Lambda p_0^2}{\Gamma}\left[\left(\zeta_1+w\frac{1-\lambda}{2}\right)\frac{q_x^2}{q^2}-\left(\zeta_2-{w}\frac{1+\lambda}{2}\right)\frac{q_y^2}{q^2}\right]
\end{equation}
%\end{widetext}
and $\xi_Q$ is the projection of $\bm{\xi}_Q$ transverse to ${\bf p}$. Only the value of $\Lambda_a$ for $q\approx q_x$ affects the equal-time angular correlator in the limit of small wavevectors. We assume that $\Lambda_a|_{q=q_x}>0$ in this paper. The concentration equation is
\begin{equation}
\partial_t \delta{c}_q=-D_cq^2\delta{c}-i v_p{c}_0p_0q_y\theta-i
\left(v_p+\frac{w}{\Lambda}\right)p_0q_x\delta{c}
\label{concpa}
\end{equation}
%\AMN{We note that at this stage it becomes clear that including a term of the form $\nabla\cdot{\bf pp}\cdot{\bf u}$ would not have qualitatively changed the concentration fluctuations. This would have led to a term $\propto q_x^2q_y/q^2\theta$ and since $q_x^2q_y/q^2<q_y$ everywhere, this would not have led to a qualitatively different anisotropy of the static structure facto of concentration fluctuations.} 
The steady state solution of \eqref{polanga} and \eqref{concpa} are simply obtained by setting $\partial_t\theta_q=\partial_t\delta c_q=0$ and solving the matrix equation
\begin{equation}
\begin{pmatrix}i  {\Lambda}_a q_x
+w\frac{q_y^2}{q^2} && i\frac{\gamma}{p_0} q_y\\i v_p{c}_0p_0q_y && i
\left(v_p+\frac{w}{\Lambda}\right)p_0q_x+D_cq^2\end{pmatrix}\begin{pmatrix}\theta_q\\\delta c_q\end{pmatrix}=\begin{pmatrix}\frac{1}{p_0}\xi_Q\\0\end{pmatrix}
\end{equation}
The solution of this equation then straightforwardly yields the disorder-averaged correlation functions for $\theta_q$:
%\begin{widetext}
\begin{equation}
\langle|\theta_q|^2\rangle=\frac{T_Qq^4[D_c^2q^4\Lambda^2+p_0^2q_x^2(w+\upsilon_p\Lambda)^2]}{-2 c_0p_0q^4q_x^2q_y^2\upsilon_p\gamma\Lambda(w+\upsilon_p\Lambda)\Lambda_a+p_0^2q_x^2(w+\upsilon_p\Lambda)^2(q_y^4w^2+q^4q_x^2\Lambda_a^2)+q^4\Lambda^2[q_y^4(D_cw+c_0\upsilon_p\gamma)^2+D_c^2q^4q_x^2\Lambda_a^2]}
\end{equation}
%\end{widetext}
In the $q\to 0$ limit, this then yields
\begin{equation}
\langle|\theta_q|^2\rangle=\frac{T_Q}{p_0^2[(q_y^4/q^4)w^2+q_x^2\Lambda_a^2]}\approx \frac{T_Qq^4}{p_0^2(q_y^4w^2+q_x^6\Lambda_a^2)}
\end{equation}
as discussed in the main text.

Beyond the linear theory, the nonlinear equation of motion for $\theta_q$ is
\begin{multline}
\partial_t\theta_q=-i\frac{\gamma}{p_0} q_y\delta{c}-i  {\Lambda}_a q_x\theta+
w\bigg[\frac{q_y}{q_x}\left(-\frac{q_y}{q_x}\theta_q+\frac{\theta_k\theta_{q-k}}{2}\right) -\theta_{q-k}\left(-\frac{k_y}{k_x}\theta_k+\frac{\theta_{k-m}\theta_m}{2}\right)\bigg]\\-i\lambda_1\delta c_{q-k}k_x\theta_k-i\lambda_2q_x\theta_{q-k}\theta_k-i\gamma_1q_y\delta c_{q-k}\delta c_k-i\gamma_2\theta_{q-k}k_x\delta c_k+\frac{1}{p_0}\xi_Q
\end{multline}
where we have used the fact that in the regime in which angular fluctuations are the largest, $q_y\sim q_x^{3/2}$ and therefore, $q^2\approx q_x^2$ for small wavevectors. Here, $\lambda_1$, $\lambda_2$, $\gamma_1$, $\gamma_2$ are coefficients that may scale independently under a renormalisation group treatment, unlike the nonlinearities with the coefficient $w$, in the square brackets, which, due to rotation invariance, must all scale the same way to preserve the form of the term in the square brackets. They will be anisotropic, just like $\Lambda_a$ but since, as discussed in the main text, all of them turn out to be irrelevant, we have not shown their (somewhat cumbersome) anisotropic forms nor expressed them in terms of coefficients introduced in \eqref{p2d} and \eqref{v2d}. However, it is easy to see the origin of these nonlinearties: $\lambda_2$, within this theory, should arise from the same terms that contribute to $\Lambda_a$ and therefore, should be equal to $\Lambda_a$. The concentration nonlinearities can be understood as arising from the concentration dependence of $\gamma$, $\lambda_a$, $\zeta_1$, $\zeta_2$, $\upsilon$. For instance, since every term in $\Lambda_a$ can be a function of concentration, $\lambda_1$ should be $\partial\Lambda_a/\partial c|_{c=c_0}$ plus additional contributions from the projection of ${\bf p}$ orthogonal to itself to obtain an equation in terms of $\theta_q$. $\gamma_1$ arises due to the concentration dependence of $\gamma$ while $\gamma_2$ appears due to the projection of ${\bf p}$ transverse to itself.

The nonlinear concentration equation is 
\begin{equation}
\partial_t \delta c_q=-D_cq^2\delta{c}-i v_p{c}_0p_0q_y\theta-i
\left(v_p+\frac{w}{\Lambda}\right)p_0q_x\delta{c}-i\lambda_c q_y\theta_{q-k}\delta c_k+iw_1q_x\delta c_{q-k}\delta c_k+iw_2q_x\theta_{q-k}\theta_k
\end{equation}
The nonlinear coefficients $\lambda_c$ and $w_1$ can rescale independently under a renormalisation group treatment but $w_2$ must scale the same way as $v_pc_0p_0$. It is simple to see the origin of these nonlinearities as well: $\lambda_c$ should be $v_p/c_0$, $w_1$ appears from a concentration dependence of $v_p$ or $\upsilon$, and $w_2$ appears from expanding $\nabla\cdot{\bf p}$ to second order in $\theta$ and therefore should have a coefficient $\propto v_pc_0p_0$.

%\section{Appendix 2}
%
%\begin{equation}
%\label{angapp}
%\partial_t\theta_q=-q^2\left[\frac{\Delta\mu \{q^2-\lambda(q_x^2-q_y^2)\}\{\zeta_2q^2-\zeta_1(q_x^2-q_y^2)\}}{2\Gamma q^4}+\Gamma_\theta K\right]\theta_q+q_xq_y{\Gamma_\theta\gamma}\delta c_q+\xi_Q=-q^2\bar{K}\theta_q+q_xq_y\Gamma_\theta\gamma\delta c_q+\xi_Q
%\end{equation}
%
%\begin{equation}
%\label{concq}
%\partial_t\delta c_q=-q^2D_c\delta c_q-\zeta_c\Delta\mu q_xq_y\theta_q
%\end{equation}
%
%\begin{equation}
%\begin{pmatrix}\bar{K}q^2 && -q_xq_y\Gamma_\theta\gamma\\\zeta_c\Delta\mu q_xq_y && D_c q^2\end{pmatrix}\begin{pmatrix}\theta_q\\\delta c_q\end{pmatrix}=\begin{pmatrix}\xi_Q\\0\end{pmatrix}
%\end{equation}
%
%\begin{equation}
%\label{apolangfluc2}
%\langle|\theta_q|^2\rangle=\frac{T_QD_c^2q^4}{(D_c\bar{K}q^4+q_x^2q_y^2\gamma\Gamma_\theta\Delta\mu\zeta_c)^2}\propto\frac{1}{q^4}.
%\end{equation}
%
%\begin{equation}
%\label{apolconcfluc}
%\langle|\delta c_q|^2\rangle=\frac{T_Qq_x^2q_y^2\Delta\mu^2\zeta_c^2}{(D_c\bar{K}q^4+q_x^2q_y^2\gamma\Gamma_\theta\Delta\mu\zeta_c)^2}
%\end{equation}
%
%\begin{equation}
%\langle\delta c({\bf x})^2\rangle=\int_{|{\bf q}|\geq1/\xi_{IM}}\frac{d{\bf q}}{4\pi^2}\langle|\delta c_q|^2\rangle\sim\xi_{IM}^2
%\end{equation}

\end{document}